\documentclass[12pt,titlepage]{article}

\def\Re{{\rm I\!R}} 
  
\newcommand{\p}{\partial} \newcommand{\xn}{\mbox{$x_{1},\ldots,x_{n}\ 
    $}} \newcommand{\J}{\left|\frac{\p(h)}{\p(z')}\right|}
\newcommand{\qed}{\mbox{{\bf Q.E.D.}}}

\newcommand{\diag}{\mbox{diag}}

\newcommand{\pd}[2]{\frac{\partial#1}{\partial#2}}

\newtheorem{theorem}{Theorem} \newtheorem{assume}{Assumption}
\newtheorem{lemma}{Lemma}

\title{Optimal Recovery of Local Truth} 
\author{Carlos C. Rodr{\'{\i}}guez \\
  Department of Mathematics and Statistics \\
    University at Albany, SUNY \\
    Albany NY 12222, USA \\
    carlos@math.albany.edu \\
      http://omega.albany.edu:8008/
}

\date{}
\begin{titlepage}
\end{titlepage}

\begin{document}
\maketitle
\tableofcontents

\begin{abstract}
  
  Probability mass curves the data space with horizons!.  Let $f$ be a
  multivariate probability density function with continuous second
  order partial derivatives.  Consider the problem of estimating the
  true value of $f(z) > 0$ at a single point $z$, from $n$ independent
  observations.  It is shown that, the fastest possible estimators
  (like the k-nearest neighbor and kernel) have minimum asymptotic
  mean square errors when the space of observations is thought as
  conformally curved. The optimal metric is shown to be generated by
  the Hessian of $f$ in the regions where the Hessian is definite.
  Thus, the peaks and valleys of $f$ are surrounded by singular
  horizons when the Hessian changes signature from Riemannian to
  pseudo-Riemannian. Adaptive estimators based on the optimal variable
  metric show considerable theoretical and practical improvements over
  traditional methods. The formulas simplify dramatically when the
  dimension of the data space is 4.  The similarities with General
  Relativity are striking but possibly illusory at this point.
  However, these results suggest that nonparametric density estimation
  may have something new to say about current physical theory.
\end{abstract}

\section{Introduction}

During the past thirty years the theory of Nonparametrics has been
dominating the scene in mathematical statistics. Parallel to the
accelerating discovery of new technical results, a consensus has been
growing among some researchers in the area, that we may be witnessing a
promising solid road towards the elusive Universal Learning Machine
(see e.g. \cite{vapnik98,devroye-gyorfi-lugosi96}).

The queen of nonparametrics is density estimation. All the fundamental
ideas for solving the new problems of statistical estimation in
functional spaces (smoothing, generalization, optimal minimax rates,
etc.) already appear in the problem of estimating the probability
density (i.e. the model) from the observed data. More over, it is now
well known that a solution for the density estimation problem
automatically implies solutions for the problems of pattern
recognition and nonparametric regression as well as for most problems
that can be expressed as a functional of the density.

In this paper I present a technical result, about optimal
nonparametric density estimation, that shows at least at a formal
level, a surprising similarity between nonparametrics and General
Relativity.  Simply put,
\begin{center}
  {\em probability mass curves the data space with horizons.}
\end{center}

What exactly it is meant by this is the subject of this paper but
before proceeding further a few comments are in order.  First of all,
let us assume that we have a set $\{x_{1},\ldots,x_{n}\}$ of data.
Each observation $x_{j}$ consisting of $p$ measurements that are
thought as the $p$ coordinates of a vector in $\Re^{p}$. To make the
data space into a probability space we endow $\Re^{p}$ with the field
of Borelians but nothing beyond that. In particular no a priori metric
structure on the data space is assumed. The $n$ observations are
assumed to be $n$ independent realizations of a given probability
measure $P$ on $\Re^{p}$. By the Lebesgue decomposition theorem, for
every Borel set $B$ we can write,

\begin{equation}
P(B) = \int_{B}f(x) \lambda(dx) + \nu(B)        \label{eq:ldt}
\end{equation}
where $\nu$ is the singular part of $P$ that assigns positive
probability mass to Borel sets of zero Lebesgue volume. Due to the
existence of pathologies like the Cantor set in one dimension and its
analogies in higher dimensions, the singular part $\nu$ cannot be
empirically estimated (see e.g. \cite{devroye-gyorfi90}). Practically
all of the papers on density estimation rule out the singular part of
$P$ a priori. The elimination of singularities by fiat has permitted
the construction of a rich mathematical theory for density estimation,
but it has also ruled out a priori the study of models of mixed
dimensionality (see \cite{renyi59}) that may be necessary for
understanding point masses and spacetime singularities coexisting with
absolutely continuous distributions.

We assume further that in the regions where $f(x)>0$ the density $f$
is of class ${\mathcal{C}}^{2}$ i.e., it has continuous second order partial
derivatives.

\subsection{Nonparametrics with the World in Mind}

The road from Classical Newtonian Physics to the physics of today can
be seen as a path paved by an increasing use of fundamental concepts
that are statistical in nature. This is obvious for statistical
mechanics, becoming clearer for quantum theory, and appearing almost
as a shock in General Relativity. Not surprisingly there have been
several attempts to take this trend further (see e.g.
\cite{jaynes57,frieden98,rodriguez98b,caticha00}) in the direction of
{\em Physics as Inference}.

Now suppose for a moment that in fact some kind of restatement of the
foundations of physics in terms of information and statistical
inference will eventually end up providing a way to advance our
current understanding of nature. As of today, that is either already a
solid fact or remains a wild speculation, depending on who you ask.
In any case, for the trend to take over, it will have to be able to
reproduce all the successes of current science and make new correct
predictions. In particular it would have to reproduce General
Relativity.  Recall that the main lesson of General Relativity is that
space and time are not just a passive stage on top of which the
universe evolves. General Relativity is the theory that tells (through
the field equation) how to build the stage (left hand side of the
equation) from the data (right hand side of the equation). The
statistical theory that tells how to build the stage of inference (the
probabilistic model) from the observed data is: {\em Nonparametric
  Density Estimation}. It is therefore reassuring to find typical
signatures of General Relativity in density estimation as this paper
does. Perhaps Physics is not just a special case of statistical
inference and all these are only coincidences of no more relevance
than for example the fact that multiplication or the logarithmic
function appear everywhere all the time. That may be so, but even in
that case I believe it is worth noticing the connection for
undoubtedly GR and density estimation have a common goal: {\em The
  dynamic building of the stage}.

More formally. Let $f$ be a multivariate probability density function
with continuous second order partial derivatives.  Consider the
problem of estimating the true value of $f(z) > 0$ at a single point
$z$, from $n$ independent observations.  It is shown that, fastest
possible estimators (including the k-nearest neighbor and kernel as
well as the rich class of estimators in
\cite[theorem3.1]{rodriguez86}) have minimum asymptotic mean square
errors when the space of observations is thought as conformally
curved. The optimal metric is shown to be generated by the Hessian of
$f$ in the regions where the Hessian is definite.  Thus, the peaks and
valleys of $f$ are surrounded by horizons where the Hessian changes
signature from Riemannian to pseudo-Riemannian.

The result for the case of generalized k-nearest neighbor estimators
\cite{rodriguez86} has circulated since 1988 in the form of a
technical report \cite{rodriguez88}. Recently I found that a special
case of this theorem has been known since 1972
\cite{fukunaga-hostetler72} and undergone continuous development in
the Pattern Recognition literature, (see e.g.
\cite{short-fukunaga81,fukunaga-flick84,fukunaga-hummels87,myles-hand90}).

\section{Estimating Densities from Data}
\label{sec:estimating}

The canonical problem of density estimation at a point $z\in \Re^{p}$
can be stated as follows: {\em Estimate $f(z)>0$ from $n$ independent
  observations of a random variable with density $f$. }

The most successful estimators of $f(z)$ attempt to approximate the
density of probability at $z$ by using the observations \xn to build
both, a small volume around $z$ and, a weight for this volume in terms
of probability mass. The density is then computed as the ratio of the
estimated mass over the estimated volume. The two classical examples
are the k-nearest neighbor (knn) and the kernel estimators.

\subsection{The knn}
The simplest and historically the first example of a nonparametric
density estimator is \cite{fix-hodges51} the knn.  The knn estimator
of $f(z)$ is defined for $k\in \{1,2,\ldots,n\}$ as,
\begin{equation}
h_{n}(z) = \frac{k/n}{\lambda_{k}}  \label{eq:hnz}
\end{equation}
where $\lambda_{k}$ is the volume of the sphere centered at the point
$z\in \Re^{p}$ of radius $R(k)$ given by the distance from $z$ to the
kth-nearest neighbor observation. If $\lambda$ denotes the Lebesgue
measure on $\Re^{p}$ we have,

\begin{equation}
\lambda_{k} = \lambda(S(R(k))) \label{eq:lk}
\end{equation}
where,
\begin{equation}
S(r) = \{ x\in \Re^{p} : \|x-z\| \le r \} \label{eq:Sr}
\end{equation}
The sphere $S(r)$ and the radius $R(k)$ are defined relative to a
given norm, $\|\cdot\|$ in $\Re^{p}$. The stochastic behavior of the
knn depends on the specific value of the integer $k$ chosen in
(\ref{eq:hnz}). Clearly, in order to achieve consistency (e.g.
stochastic convergence of $h_{n}(z)$ as $n\rightarrow\infty$ towards
the true value of $f(z)>0$) it is necessary to choose $k=k(n)$ as a
function of $n$. The volumes $\lambda_{k}$ must shrink, to control the
bias, and consequently we must have $k/n\rightarrow 0$ for $h_{n}(z)$
to be able to approach a strictly positive number. On the other hand,
we must have $k\rightarrow\infty$ to make the estimator dependent on
an increasing number $k$ of observations and in this way to control
its variance.  Thus, for the knn to be consistent, we need $k$ to
increase with $n$ but at a rate slower than $n$ itself.

The knn estimator depends not only on $k$ but also on a choice of
norm. The main result of this paper follows from the characterization
of the $\|\cdot\|$ that, under some regularity conditions, produces
the best asymptotic (as $n\rightarrow\infty$) performance for density
estimators.

\subsection{The kernel}
If we consider only regular norms $\|\cdot\|$, in the sense that for
all sufficiently small values of $r>0$,
\begin{equation}
\lambda(S(r)) = \lambda(S(1)) r^{p} \equiv \beta r^{p} \label{eq:SrS1}
\end{equation}
then, the classical kernel density estimator can be written as:

\begin{equation}
g_{n}(z) = \frac{M_{\mu}}{\lambda(S(\mu))} \label{eq:gnz}
\end{equation}
where,

\begin{equation}
M_{\mu} = \frac{1}{n}\left(\sum_{x_{j}\in S(\mu)} K_{\mu^{-1}}(x_{j}-z)\right) \label{eq:Mmu}
\end{equation}
The smoothing parameter $\mu = \mu(n)$ is such that $k=[n\mu^{p}]$
satisfies the conditions for consistency of the knn, $K_{\mu^{-1}}(x)
= K(\mu^{-1}x)$ where the kernel function $K$ is a non negative
bounded function with support on the unit sphere (i.e. $K(x)=0$ for
$\|x\| > 1$) and satisfying,

\begin{equation}
\int_{\|x\|\le 1} K(x) dx = \beta       \label{eq:S1Kx}
\end{equation}
Notice that for the constant kernel (i.e. $K(x)=1$ for $\|x\|\le 1$)
the estimator (\ref{eq:gnz}) approximates $f(z)$ by the proportion of
observations inside $S(\mu)$ over the volume of $S(\mu)$. The general
kernel function $K$ acts as a weight function allocating different
weights $K_{\mu^{-1}}(x_{j}-z)$ to the $x_{j}$'s inside $S(\mu)$. To
control bias (see (\ref{eq:Egn3}) below) the kernel $K$ is usually
taken as a decreasing radially symmetric function in the metric
generated by the norm $\|\cdot\|$. Thus, $K_{\mu^{-1}}(x_{j}-z)$
assigns a weight to $x_{j}$ that decreases with its distance to $z$.
This has intuitive appeal, for the observations that lie closer to $z$
are less likely to fall off the sphere $S(\mu)$, under repeated
sampling, than the observations that are close to the boundary of
$S(\mu)$.

The performance of the kernel as an estimator for $f(z)$ depends first
and foremost on the value of the smoothness parameter $\mu$. The
numerator and the denominator of $g_{n}(z)$ depend not only on $\mu$
but also on the norm $\|\cdot\|$ chosen and the form of the kernel
function $K$. As it is shown in theorem (\ref{th:changeA}) these three
parameters are inter-related.

\subsection{Double Smoothing Estimators}

The knn (\ref{eq:hnz}) and the kernel (\ref{eq:gnz}) methods are two
extremes of a continuum. Both, $h_{n}(z)$ and $g_{n}(z)$ estimate
$f(z)$ as {\em probability-mass-per-unit-volume}. The knn fixes the
mass to the deterministic value $k/n$ and lets the volume
$\lambda_{k}$ to be stochastic, while the kernel method fixes the
volume $\lambda(S(\mu))$ and lets the mass $M_{\mu}$ to be random.
The continuum gap between (\ref{eq:hnz}) and (\ref{eq:gnz}) is filled
up by estimators that stochastically estimate mass and volume by
smoothing the contribution of each sample point with different
smoothing functions for the numerator and denominator (see
\cite{rodriguez86}).

Let $b\ge 1$ and assume, without loss of generality that $bk$ is an
integer. The double smoothing estimators with deterministic weights
are defined as,
\begin{equation}
  \label{eq:doublesmooth}
  f_{n}(z) = \frac{\frac{1}{n}\sum_{i=1}^{n} 
    K\left(\frac{z-x_{i}}{R(k)}\right)}
    {\frac{1}{cb} \sum_{i=1}^{bk} \omega_{i} \lambda_{i} }
\end{equation}
where,
\begin{equation}
  \label{eq:wi}
  \omega_{i} = \int_{(i-1)/bk}^{i/bk} \omega(u) du
\end{equation}
and $\omega(\cdot)$ is a probability density on $[0,1]$ with mean $c$.

\section{The Truth as $n\rightarrow\infty$ ?}
\label{section:thetruth}

In nonparametric statistics, in order to assess the quality of an
estimator $f_{n}(z)$ as an estimate for $f(z)$, it is necessary to
choose a criterion for judging how far away is the estimator from what
it tries to estimate. This is sometimes regarded as revolting and
morally wrong by some Bayesian Fundamentalists. For once you choose a
loss function and a prior, logic alone provides you with the Bayes
estimator and the criterion for judging its quality. That is
desirable, but there is a problem in high dimensional spaces. In
infinite dimensional hypothesis spaces (i.e. in nonparametric
problems) almost all priors will convince you of the wrong thing! (see
e.g.  \cite{lecam-yang90,diaconnis-freedman86} for a non-regular way
out see \cite{rodriguez97}). These kind of Bayesian nonparametric
results provide a mathematical proof that: {\em almost all fundamental
  religions are wrong,} (more data can only make the believers more
sure that the wrong thing is true!). An immediate corollary is that:
{\em Subjective Bayesians can't go to Heaven}. Besides, the choice of
goodness of fit criterion is as ad-hoc (an equivalent) to the choice
of a loss function.

\subsection{The Natural Invariant Loss Function and Why the MSE is not
  that Bad}

The most widely studied goodness of fit criterion is the Mean Square
Error (MSE) defined by,

\begin{equation}
(\mbox{MSE}) = E|f_{n}(z) - f(z)|^{2}   \label{eq:mse}
\end{equation}
where the expectation is over the joint distribution of the sample
\xn. By adding and subtracting $T=Ef_{n}(z)$ and expanding the square,
we can express the MSE in the computationally convenient form,

\begin{eqnarray}
(\mbox{MSE}) &=& E|f_{n}(z) - T|^{2} + |T - f(z)|^{2} \nonumber \\
        &=&  \mbox{ (variance) $ + \mbox{(bias)}^{2}$ } \label{eq:mse2}
\end{eqnarray}
By integrating (\ref{eq:mse}) over the $z\in \Re^{p}$ and
interchanging $E$ and $\int$ (OK by Fubbini's theorem since the
integrand $\ge 0$) we obtain,

\begin{equation}
(\mbox{MISE}) = E\int |f_{n}(z) - f(z)|^{2} dz  \label{eq:mise}
\end{equation}
The Mean Integrated Square Error (MISE) is just the expected $L^{2}$
distance of $f_{n}$ from $f$. Goodness of fit measures based on the
$(MSE)$ have two main advantages: They are often easy to compute and
they enable the rich Hilbertian geometry of $L^{2}$. On the other hand
the $(MSE)$ is unnatural and undesirable for two reasons: Firstly, the
$(MSE)$ is only defined for densities in $L^{2}$ and this rules out
a priori all the densities in $L^{1}\setminus L^{2}$ which is
unacceptable. Secondly, even when the $(MISE)$ exists, it is difficult
to interpret (as a measure of distance between densities) due to its
lack of invariance under relabels of the data space. Many researchers
see the expected $L^{1}$ distance between densities as the natural
loss function in density estimation. The $L^{1}$ distance does in fact
exist for all densities and it is easy to interpret but it lacks the
rich geometry generated by the availability of the inner product in
$L^{2}$. A clean way out is to use the expected $L^{2}$
distance between the wave functions $\psi_{n}=\sqrt{f_{n}}$ and
$\psi=\sqrt{f}$.

\begin{theorem}
\label{th:wavenorm}
The $L^{2}$ norm of wave functions is invariant under relabels of the
data space, i.e.,

\begin{equation}
\int |\psi_{n}(z) - \psi(z)|^{2} dz = 
                \int |\tilde{\psi}_{n}(z') - \tilde{\psi}(z')|^{2} dz' \label{eq:psin}
\end{equation}
where $z = h(z')$ with $h$ any one-to-one smooth function.
\end{theorem}
{\bf Proof:} Just change the variables. From, the change of variables
theorem the pdf of $z'$ is,

\begin{equation}
\tilde{f}(z') = f(h(z')) \J     \label{eq:tildef}
\end{equation}
from where the wave function of $z'$ is given by,

\begin{equation}
\tilde{\psi}(z') = \psi(h(z')) \J^{1/2} \label{eq:tildepsi}
\end{equation}
Thus, making the substitution $z=h(z')$ we get,

\begin{eqnarray}
\int |\psi_{n} - \psi|^{2} dz &=& 
        \int |\psi_{n}(h(z')) - \psi(h(z'))|^{2} \J dz' \nonumber \\
        &=& \int |\psi_{n}\J^{1/2} - \psi\J^{1/2}|^{2} dz' \nonumber \\
        &=& \int |\tilde{\psi}_{n} - \tilde{\psi}|^{2} dz'
\end{eqnarray}
\qed

The following theorem shows that a transformation of the MSE of a
consistent estimator provides an estimate for the expected $L^{2}$
norm between wave functions.

\begin{theorem}
\label{th:wavemse}
Let $f_{n}(z)$ be a consistent estimator of $f(z)$. Then,

\begin{equation}
E \int |\psi_{n} - \psi|^{2} dz = 
        \frac{1}{4} \int \frac{E|f_{n}(z) - f(z)|^{2}}{f(z)} dz 
        + \mbox{(smaller order terms)}          \label{eq:Epsi}
\end{equation}
\end{theorem}
{\bf Proof:} A first order Taylor expansion of $\sqrt{x}$ about
$x_{0}$ gives,

\begin{equation}
\sqrt{x} - \sqrt{x_{0}} = \frac{1}{2}\frac{(x-x_{0})}{\sqrt{x_{0}}} + o((x-x_{0})^{2})
                        \label{eq:sqrtx}
\end{equation}
Substituting $x=f_{n}(z), x_{0}=f(z)$ into (\ref{eq:sqrtx}) squaring
both sides and taking expectations we obtain,

\begin{equation}
E|\psi_{n}(z) - \psi(z)|^{2} = \frac{1}{4}\frac{E|f_{n}(z)-f(z)|^{2}}{f(z)}
        + o(E|f_{n}(z)-f(z)|^{2})       \label{eq:Epsin}
\end{equation}
integrating over $z$ and interchanging $E$ and $\int$ we arrive at
(\ref{eq:Epsi}).
\\
\qed

Proceeding as in the proof of theorem~\ref{th:wavenorm} we can show
that

\begin{equation}
\int \frac{|f_{n} -f|^{2}}{f} dz = 
        \int \frac{|\tilde{f}_{n} -\tilde{f}|^{2}}{\tilde{f}} dz' \label{eq:intfn}
\end{equation}
where, as before, $z\leftrightarrow z'$ is any one-to-one smooth
transformation of the data space and $\tilde{f}$ is the density of
$z'$. Thus, it follows from (\ref{eq:intfn}) that the leading term on
the right hand side of (\ref{eq:Epsi}) is also invariant under
relabels of the data space. The nice thing about the $L^{2}$ norm of
wave functions, unlike (\ref{eq:intfn}), is that it is defined even
when $f(z)=0$.

\section{Some Classic Asymptotic Results}
\label{section:some}

We collect here the well known Central Limit Theorems (CLT) for the
knn and kernel estimators together with some remarks about
nonparametric density estimation in general.  The notation and the
formulas introduced here will be needed for computing the main result
about optimal norms in the next section.

\begin{assume}
\label{assume:f}
Let $f$ be a pdf on $\Re^{p}$ of class ${\mathcal{C}}^{2}$ with non singular
Hessian, $H(z)$ at $z\in \Re^{p}$, and with $f(z) > 0$, i.e., the
matrix of second order partial derivatives of $f$ at $z$ exists, it is
non singular and its entries are continuous at $z$.
\end{assume}

\begin{assume}
\label{assume:K}
Let $K$ be a bounded non negative function defined on the unit sphere,
$S_{0} = \{x\in \Re^{p}: \|x\|\le 1\}$ and satisfying,
\begin{eqnarray}
\int_{\|x\|\le 1} K(x) dx = \lambda(S_{0}) \equiv \beta \label{eq:intK} \\
\int_{\|x\|\le 1} x K(x) dx = 0 \in \Re^{p}  \label{eq:intxK}
\end{eqnarray}
\end{assume}

\begin{theorem}[CLT for knn]
\label{th:cltknn}
Under assumption~\ref{assume:f}, if $k=k(n)$ is taken in the
definition of the knn (\ref{eq:hnz}) in such a way that for some $a>0$
\begin{equation}
\lim_{n\rightarrow\infty} n^{-4/(p+4)} k = a    \label{eq:limn}
\end{equation}
then, if we let $Z_{n} = \sqrt{k}(h_{n}(z) - f(z))$ we have,

\begin{equation}
\lim_{n\rightarrow\infty} P(Z_{n} \le t) = 
        \int_{-\infty}^{t} \frac{1}{\sqrt{2\pi}}
                \exp\left(-\frac{(y-B(z))^{2}}{2V(z)}\right) dy  \label{eq:limn1}
\end{equation}
where,

\begin{equation}
B(z) = \left(\frac{a^{\frac{p+4}{2p}}}{2f^{2/p}(z)}\right)
        \left\{\beta^{-1-2/p} \int_{\|x\|\le 1} x^{T}H(z)x dx \right\} \label{eq:Bzknn}
\end{equation}
and,

\begin{equation}
V(z) = f^{2}(z) \label{eq:Vzknn}
\end{equation}
\end{theorem}
{\bf Proof:} This is a special case of \cite[theorem3.1]{rodriguez86}.

\begin{theorem}[CLT for kernel]
  Under assumptions~\ref{assume:f}, and~\ref{assume:K} if $\mu=\mu(n)$
  is taken in the definition of the kernel (\ref{eq:gnz}) in such a
  way that for some $a>0$, $k=[n\mu^{p}]$ satisfies (\ref{eq:limn})
  then, if we let $Z_{n} = \sqrt{k}(g_{n}(z) - f(z))$ we have
  (\ref{eq:limn1}) where now,

\begin{equation}
B(z) = \left(\frac{a^{\frac{p+4}{2p}}}{2}\right)
        \left\{\beta^{-1} \int_{\|x\|\le 1} x^{T}H(z)x K(x) dx \right\} \label{eq:Bzkernel}
\end{equation}
and,

\begin{equation}
V(z) = f(z) \left\{ \beta^{-2} \int_{\|x\|\le 1} K^{2}(x) dx \right\} \label{eq:Vzkernel}
\end{equation}

\end{theorem}
{\bf Proof:} The sample \xn is assumed to be iid $f$ and therefore the
kernel estimator $g_{n}(z)$ given by (\ref{eq:gnz}) and (\ref{eq:Mmu})
is a sum of iid random variables. Thus, the classic CLT applies and we
only need to verify the rate (\ref{eq:limn}) and the asymptotic
expressions for the bias (\ref{eq:Bzkernel}) and variance
(\ref{eq:Vzkernel}). We have,

\begin{eqnarray}
E\left[ g_{n}(z) \right] &=& \frac{1}{\beta \mu^{p}}\frac{1}{n}
        \sum_{j=1}^{n}\int K\left(\frac{x_{j}-z}{\mu}\right)f(x_{j}) dx_{j} \label{eq:Egn1} \\
&=& \frac{1}{\beta\mu^{p}}\int K(y) f(z+\mu y)\mu^{p} dy        \label{eq:Egn2} \\
&=& \int\frac{K(y)}{\beta}\left\{f(z) + \mu\nabla f(z)\cdot y + \frac{\mu^{2}}{2}y^{T}H(z)y 
        + o(\mu^{2})\right\} dy         \label{eq:Egn3} \\
&=& f(z) + \frac{\mu^{2}}{2\beta}\int y^{T}H(z)y K(y) dy + o(\mu^{2}) \label{eq:Egn4}
\end{eqnarray}
where we have changed the variables of integration to get
(\ref{eq:Egn2}), used assumption~\ref{assume:f} and Taylor's theorem
to get (\ref{eq:Egn3}) and used assumption~\ref{assume:K} to obtain
(\ref{eq:Egn4}). For the variance we have,

\begin{eqnarray}
\mbox{var}(g_{n}(z)) &=& \frac{1}{n\beta^{2}\mu^{2p}}\mbox{var}
        \left(K((X-z)/\mu)\right) \label{eq:varfn1} \\
&=& \frac{1}{n\beta^{2}\mu^{2p}}\left\{ \int_{\|y\|\le 1} f(z+\mu y) K^{2}(y) \mu^{p}dy\right.\nonumber \\
& &     - \left. \left(\int_{\|y\|\le 1}f(z+\mu y)K(y) \mu^{p}dy\right)^{2} \right\} \label{eq:varfn2} \\
&=& \frac{f(z)}{n\beta^{2}\mu^{p}}\int_{\|y\|\le 1}K^{2}(y)dy + o\left(\frac{1}{n\mu^{p}}\right)
                \label{eq:varfn3}
\end{eqnarray}
where we have used $\mbox{var}(K)=EK^{2}-(EK)^{2}$ and changed the
variables of integration to get (\ref{eq:varfn2}), used
assumption~\ref{assume:f} and (0th order) Taylor's theorem to get
(\ref{eq:varfn3}). Hence, the theorem follows from (\ref{eq:Egn4}) and
(\ref{eq:varfn3}) after noticing that (\ref{eq:limn}) and $k=n\mu^{p}$
imply,

\begin{eqnarray}
\sqrt{k}\mu^{2} &=& k^{\frac{4+p}{2p}}n^{-2/p} = (n^{-\frac{4}{p+4}}k)^{\frac{p+4}{2p}}
        \longrightarrow a^{\frac{p+4}{2p}} \label{eq:sqrtkmu} \\
\frac{k}{n\mu^{p}} &=& \frac{k}{k} = 1  \label{eq:knmu}
\end{eqnarray}
\qed

\begin{theorem}[CLT for double smoothers]
Consider the estimator $f_{n}(z)$ defined in (\ref{eq:doublesmooth}).
  Under assumptions~\ref{assume:f}, \ref{assume:K}, and (\ref{eq:limn})
  if we let $Z_{n} = \sqrt{k}(f_{n}(z) - f(z))$ we have
  (\ref{eq:limn1}) where now,

\begin{equation}
B(z) = \left(\frac{a^{\frac{p+4}{2p}}}{2[\beta f(z)]^{2/p}}\right)
        \beta^{-1} \left\{\int_{\|x\|\le 1} x^{T}H(z)x 
          \left[K(x)+\lambda_{0}\right] dx \right\} \label{eq:Bzfn}
\end{equation}
and,

\begin{equation}
V(z) = f^{2}(z) \left\{ \beta^{-1} \int_{\|x\|\le 1} 
          K^{2}(x) dx - \lambda_{1}\right\} \label{eq:Vzfn}
\end{equation}
with,
\begin{eqnarray}
  \label{eq:lambda0}
  \lambda_{0} &=& \frac{b^{2/p}}{c}\int_{0}^{1}u^{1+\frac{2}{p}}
                        \omega(u) du - 1 \\
  \label{eq:lambda1}
  \lambda_{1} &=& 1 - c^{-2}b^{-1} \int_{0}^{1}\left\{\int_{y}^{1}
                  \omega(x) dx \right\}^{2} dy
\end{eqnarray}
\end{theorem}
{\bf Proof:} See \cite[theorem3.1]{rodriguez86}. Remember to
substitute $K$ by $\beta^{-1}K$ since in the reference the Kernels are
probability densities and in here we take them as weight functions
that integrate to $\beta$.

\subsection{Asymptotic Mean Square Errors}

Let $f_{n}$ be an arbitrary density estimator and let
$Z_{n}=\sqrt{k}(f_{n}(z)-f(z))$. Now suppose that $f_{n}(z)$ is
asymptotically normal, in the sense that when $k=k(n)$ satisfies
(\ref{eq:limn}) for some $a>0$, we have (\ref{eq:limn1}) true. Then,
all the moments of $Z_{n}$ will converge to the moments of the
asymptotic Gaussian. In particular the mean and the variance of
$Z_{n}$ will approach $B(z)$ and $V(z)$ respectively. Using,
(\ref{eq:mse2}) and (\ref{eq:limn}) we can write,

\begin{equation}
\lim_{n\rightarrow\infty} n^{4/(p+4)} E|f_{n}(z)-f(z)|^{2} =
        \frac{V(z)}{a} + \frac{B^{2}(z)}{a}     \label{eq:amse}
\end{equation}
We call the right hand side of (\ref{eq:amse}) the asymptotic mean
square error (AMSE) of the estimator $f_{n}(z)$. The value of $a$ can
be optimized to obtain a global minimum for the (AMSE) but it is well
known in nonparametrics that the rate $n^{-4/(p+4)}$ is best possible
(in a minimax sense) under the smoothness asumption~\ref{assume:f}
(see e.g. \cite{kn:brgmv.81}). We can take care of the knn, the
kernel, and the double smoothing estimators simultaneously by noticing
that in all cases,

\begin{equation}
\mbox{(AMSE)} = \alpha_{1}a^{-1} + \alpha_{2}a^{4/p} \label{eq:alphas} 
\end{equation}
has a global minimum of,

\begin{equation}
\mbox{(AMSE)}^{*} = \left\{(1+4/p)\left(\frac{p}{4}\right)^{\frac{4}{p+4}}\right\}
        \alpha_{1}^{\frac{4}{p+4}}\alpha_{2}^{\frac{p}{p+4}} \label{eq:amse*}
\end{equation}
achieved at,

\begin{equation}
a^{*} = \left(\frac{p\alpha_{1}}{4\alpha_{2}}\right)^{\frac{p}{p+4}}
                \label{eq:a*}
\end{equation}
Replacing the corresponding values for $\alpha_{1}$ and $\alpha_{2}$
for the knn, for the kernel, and for the double smoothing estimators,
we obtain that in all cases, 

\begin{equation}
\mbox{(AMSE)}^{*} = \mbox{(const. indep. of $f$)} \left\{ f(z)
\left(\frac{\Delta^{2}}{f(z)}\right)^{\frac{p}{p+4}} \right\}   \label{eq:amse*2}
\end{equation}
where,

\begin{eqnarray}
\Delta &=& \int_{\|x\|\le 1} x^{T}H(z)x G(x) dx \label{eq:Delta1} \\
       &=& \sum_{j=1}^{p} \rho_{j} \frac{\p^{2}f}{\p z_{j}^{2}}(z)  \label{eq:Delta2}
\end{eqnarray}
with $G(x)=1$ for the knn, $G(x)=K(x)$ for the kernel,
$G(x)=K(x)+\lambda_{0}$ for the double smoothers (see (\ref{eq:Bzfn})
and (\ref{eq:lambda0})) and, if $e_{j}$ denotes the $j$th canonical
basis vector (all zeroes except a 1 at position $j$),

\begin{equation}
\rho_{j} = \int_{\|x\|\le 1} (x\cdot e_{j})^{2} G(x) dx \label{eq:rhoj}
\end{equation}
Notice that (\ref{eq:Delta2}) follows from (\ref{eq:Delta1}),
(\ref{eq:intxK}) and the fact that $H(z)$ is the Hessian of $f$ at
$z$. The generality of this result shows that (\ref{eq:amse*2}) is
typical for density estimation. Thus, when $f_{n}$ is either the knn,
the kernel, or one of the estimators in (\cite[theorem3.1]
{rodriguez86}), we have:

\begin{equation}
\lim_{n\rightarrow\infty}n^{4/(p+4)}E|f_{n}(z)-f(z)|^{2} \ge c f(z)
        \left(\frac{\Delta^{2}}{f(z)}\right)^{\frac{p}{p+4}}    \label{eq:bound}
\end{equation}
The positive constant $c$ may depend on the particular estimator but
it is independent of $f$. Dividing both sides of (\ref{eq:bound}) by
$f(z)$, integrating over $z$, using theorem~\ref{th:wavemse} and
interchanging $E$ and $\int$ we obtain,

\begin{equation}
\lim_{n\rightarrow\infty}n^{4/(p+4)}E\int|\psi_{n}(z)-\psi(z)|^{2} dz \ge 
        4c \int \left|\frac{\Delta}{\psi(z)}\right|^{\frac{2p}{p+4}} dz \label{eq:bound1}
\end{equation}
The worst case scenario is obtained by the model $f=\psi^{2}$ that
maximizes the action given by the right hand side of
(\ref{eq:bound1}),

\begin{equation}
{\cal L} = \int \left| \frac{1}{\psi(z)}
         \sum_{j=1}^{p} \rho_{j} \frac{\p^{2}\psi^{2}}{\p z_{j}^{2}}(z)
        \right|^{\frac{2p}{p+4}} dz \label{eq:L}
\end{equation}
This is a hard variational problem. However, it is worth noticing that
the simplest case is obtained when the exponent is $1$, i.e. when the
dimension of the data space is $p=4$. Assuming we were able to find a
solution, this solution would still depend on the $p$ parameters
$\rho_{1},\ldots,\rho_{p}$. A choice of $\rho_{j}$'s is equivalent to
the choice of a global metric for the data space. Notice also, that
the exponent becomes $2$ for $p=\infty$ and that for $p\ge 3$ (but not
for $p=1$ or $2$) there is the possibility of non trivial (i.e.
different from uniform) super-efficient models for which estimation can
be done at rates higher than $n^{-4/(p+4)}$. These super-efficient
models are characterized as the non negative solutions of the Laplace
equation in the metric generated by the $\rho_{j}$'s, i.e., non
negative ($f(z)\ge 0$) solutions of,

\begin{equation}
\sum_{j=1}^{p} \rho_{j} \frac{\p^{2}f}{\p z_{j}^{2}}(z) = 0 \label{eq:laplace}
\end{equation}
Recall that there are no non trivial (different from constant) non
negative super-harmonic functions in dimensions one or two but there
are plenty of solutions in dimension three and higher. For example the
Newtonian potentials,

\begin{equation}
f(z) = c\|z\|_{\rho}^{-(p-2)}   \label{eq:newtonian}
\end{equation}
with the norm,

\begin{equation}
\|z\|_{\rho}^{2} = \sum_{j=1}^{p}\left(\frac{z_{j}}{\sqrt{\rho_{j}}}\right)^{2} \label{eq:normrho}
\end{equation}
will do, provided the data space is compact. The existence of (hand
picked) super-efficient models is what made necessary to consider best
rates only in the minimax sense. Even though we can estimate a
Newtonian potential model at faster than usual nonparametric rates, in
any neighborhood of the Newtonian model the worst case scenario is at
best estimated at rate $n^{-4/(p+4)}$ under second order smoothness
conditions.

\section{Choosing the Optimal Norm}

All finite ($p<\infty$) dimensional Banach spaces are isomorphic (as
Banach spaces) to $\Re^{p}$ with the euclidian norm. This means, among
other things, that in finite dimensional vector spaces all norms
generate the same topology. Hence, the spheres $\{x\in \Re^{p}:
\|x\|\le r\}$ are Borelians so they are Lebesgue measurable and thus,
estimators like the knn (\ref{eq:hnz}) are well defined for arbitrary
norms. It is possible, in principle, to consider norms that are not
coming from inner products but in this paper we look only at Hilbert
norms $\|\cdot\|_{A}$ of the form,

\begin{equation}
\|z\|_{A}^{2} = z^{T}A^{T}Az    \label{eq:normA}
\end{equation}
where $A\in\Lambda$ with $\Lambda$ defined as the open set of real
non-singular $p\times p$ matrices. For each $A\in\Lambda$ define the
unit sphere,

\begin{equation}
S_{A} = \{x\in\Re^{p}: x^{T}A^{T}Ax \le 1\}     \label{eq:SA}
\end{equation}
its volume,

\begin{equation}
\beta_{A} = \lambda(S_{A}) = \int_{S_{A}}\lambda(dx)    \label{eq:betaA}
\end{equation}
and the $A$-symmetric (i.e. $\|\cdot\|_{A}$ radially symmetric)
kernel, $K_{A}$,

\begin{equation}
K_{A}(x) = (K\circ A)(x) = K(Ax)        \label{eq:KAx}
\end{equation}
where $K$ satisfies assumption~\ref{assume:K} and it is $I$-symmetric,
i.e., radially symmetric in the euclidian norm. This means that $K(y)$
depends on $y$ only through the euclidian length of $y$, i.e. there
exists a function $F$ such that,

\begin{equation}
K(y) = F(y^{T}y)        \label{eq:KyF}
\end{equation}
The following simple theorem shows that all $A$-symmetric functions
are really of the form (\ref{eq:KAx}).

\begin{theorem}
\label{th:Asymmetry}
For any $A\in \Lambda$, $\tilde{K}$ is $A$-symmetric if and only if we
can write
\begin{equation}
\tilde{K}(x) = K(Ax) \mbox{\ for all\ } x\in\Re^{p}     \label{eq:tildeK}
\end{equation}
for some $I$-symmetric $K$.
\end{theorem}
{\bf Proof:} $\tilde{K}(x)$ is $A$-symmetric iff $\tilde{K}(x) =
F(\|x\|_{A}^{2})$ for some function $F$. Choose
$K(x)=\tilde{K}(A^{-1}x)$. This $K$ is $I$-symmetric since
$K(x)=F\left((AA^{-1}x)^{T}(AA^{-1}x)\right) = F(x^{T}x)$. More over,
$\tilde{K}(x) = \tilde{K}(A^{-1}(Ax)) = K(Ax)$. Thus,
(\ref{eq:tildeK}) is necessary for $A$-symmetry. It is also obviously
sufficient since the assumed $I$-symmetry of $K$ in (\ref{eq:tildeK})
implies that $\tilde{K}(x) = F((Ax)^{T}(Ax)) = F(\|x\|_{A}^{2})$ so it
is $A$-symmetric.\\
\qed

An important corollary of theorem~\ref{th:Asymmetry} is,

\begin{theorem}
\label{th:Asymmetry2}
Let $A,B\in \Lambda$. Then, $\tilde{K}$ is $AB$-symmetric if and only
if $\tilde{K}_{B^{-1}}$ is $A$-symmetric.
\end{theorem}
{\bf Proof:} By the first part of theorem~\ref{th:Asymmetry} we have
that $\tilde{K}=K\circ A\circ B$ with $K$ some $I$-symmetric. Thus,
$\tilde{K}\circ{B^{-1}}= K\circ A$ is $A$-symmetric by
the second part of theorem~\ref{th:Asymmetry}.\\
\qed

Let us denote by $\beta(A,K)$ the total volume that a kernel $K$
assigns to the unit $A$-sphere $S_{A}$, i.e.,

\begin{equation}
\beta(A,K) = \int_{S_{A}} K(x) dx       \label{eq:betaAK}
\end{equation}
In order to understand the effect of changing the metric on a density
estimator like the kernel (\ref{eq:gnz}), it is convenient to write
$g_{n}$ explicitly as a function of everything it depends on; The
point $z$, the metric $A$, the $A$-symmetric kernel function
$\tilde{K}$, the positive smoothness parameter $\mu$ and, the data set
$\{x_{1},\ldots,x_{n}\}$. Hence, we define,

\begin{equation}
g_{n}(z;A,\tilde{K},\mu,\{x_{1},\ldots,x_{n}\}) = 
 \frac{\frac{1}{n}\sum_{j=1}^{n} \tilde{K}\left(
                        \frac{x_{j}-z}{\mu}\right)}
        {\beta(A,\tilde{K}) \mu^{p}} \label{eq:gnzA}
\end{equation}

The following result shows that kernel estimation with metric $AB$ is
equivalent to estimation of a transformed problem with metric $A$.
The explicit form of the transformed problem indicates that a
perturbation of the metric should be regarded as composed of two
parts: Shape and volume. The shape is confounded with the form of the
kernel and the change of volume is equivalent to a change of the
smoothness parameter.

\begin{theorem}
\label{th:changeA}
Let $A,B\in\Lambda$, $\mu > 0$, and $\tilde{K}$ an $AB$-symmetric
kernel. Then, for all $z\in\Re^{p}$ and all data sets
$\{x_{1}\ldots,x_{n}\}$ we have,
\begin{equation}
g_{n}(z;AB,\tilde{K},\mu,\{x_{1},\ldots,x_{n}\}) = 
        g_{n}(\hat{B}z;A,\tilde{K}\circ B^{-1},|B|^{-1/p}\mu,
                \{\hat{B}x_{1},\ldots,\hat{B}x_{n}\})  \label{eq:gnzAB}
\end{equation}
where $|B|$ denotes the determinant of $B$ and $\hat{B}=|B|^{-1/p}B$
is the matrix $B$ re-scaled to have unit determinant.
\end{theorem}
{\bf Proof:} To simplify the notation let us denote,

\begin{equation}
\mu_{B} = \frac{\mu}{|B|^{1/p}} \label{eq:muB}
\end{equation}
Substituting $AB$ for $A$ in (\ref{eq:gnzA}) and using
theorem~\ref{th:Asymmetry} we can write the left hand side of
(\ref{eq:gnzAB}) as,

\[
\frac{\frac{1}{n}\sum_{j=1}^{n} K\left(
    AB\left(\frac{x_{j}-z}{\mu}\right)\right)} {\beta(AB,\tilde{K})
  \mu^{p}} = \frac{\frac{1}{n}\sum_{j=1}^{n} (K\circ A)\left(
    \frac{\hat{B}x_{j}-\hat{B}z}{\mu_{B}}\right)} {\beta(A,K\circ A)
  (\mu_{B})^{p}}
\]
where $K$ is some $I$-symmetric kernel and we have made the change of
variables $x=B^{-1}y$ in $\beta(AB,\tilde{K})$ (see (\ref{eq:betaAK})
). The last expression is the right hand side of (\ref{eq:gnzAB}).
Notice that, $K\circ A = \tilde{K}_{B^{-1}}$ is
$A$-symmetric.\\
\qed

\subsection{Generalized knn Case with Uniform Kernel}
\label{sec:gknnc}

In this section we find the norm of type (\ref{eq:normA}) that
minimizes (\ref{eq:amse*2}) for the estimators of the knn type with
uniform kernel which include the double smoothers with $K(x)=1$. As it
is shown in theorem \ref{th:changeA} a change in the determinant of
the matrix that defines the norm is equivalent to changing the
smoothness parameter. The quantity (\ref{eq:normA}) to be minimized
was obtained by fixing the value of the smoothness parameter to the
best possible, i.e. the one that minimizes the expression of the
(AMSE) (\ref{eq:amse}). Thus, to search for the best norm at a fix
value of the smoothness parameter we need to keep the determinant of
the matrix that defines the norm constant. We have,
\begin{theorem}
  \label{th:optA}
Consider the problem,
\begin{equation}
  \label{eq:minA}
  \min_{|A| = 1} \left(\int_{S_{A}} x^{T}H(z)x dx\right)^{2}
\end{equation}
where the minimum is taken over all \(p\times p\) matrices with
determinant one, \(S_{A}\) is the unit \(A\)-ball and \(H(z)\) is the
Hessian of the density \(f\in {\mathcal{C}}^{2}\) at \(z\) which is
assumed to be nonsingular. 

When $H(z)$ is indefinite, i.e. when $H(z)$ has both positive and
negative eigenvalues, the objective function in (\ref{eq:minA})
achieves its absolute minimum value of zero when $A$ is taken as,
\begin{equation}
  \label{eq:bestA0}
  A = c^{-1} \diag (\sqrt{\frac{\xi_{1}}{p-m}},\ldots,
           \sqrt{\frac{\xi_{m}}{p-m}},\sqrt{\frac{\xi_{m+1}}{m}},\ldots,
             \sqrt{\frac{\xi_{p}}{m}}) M
\end{equation}
where the $\xi_{j}$ are the absolute value of the eigenvalues of
$H(z)$, $m$ is the number of positive eigenvalues, $M$ is the matrix of
eigenvectors and $c$ is a normalization constant to get $\det A = 1$
(see the proof for more detailed definitions).

If $H(z)$ is definite, i.e. when $H(z)$ is either positive or
negative definite, then for all $p\times p$ real matrices $A$ with
$\det A = 1$ we have,
\begin{equation}
  \label{eq:lb4obj}
  \left| \int_{S_{A}} x^{T}H(z)x dx \right| \ge
    \frac{2^{p}\pi}{p+3}\  p \left| \det H(z) \right|^{1/p}
\end{equation}
with equality if and only if,
\begin{equation}
  \label{eq:bestA2}
  A = \frac{H_{+}^{1/2}(z)}{|H_{+}^{1/2}(z)|^{1/p}} 
\end{equation}
\end{theorem}
where $H_{+}^{1/2}(z)$ denotes the positive definite square-root of
$H(z)$ (see the proof below for explicit definitions).

\textbf{Proof:} Since \(f\in {\mathcal{C}}^{2}\) the Hessian is a real
symmetric matrix and we can therefore find an orthogonal matrix \(M\)
that diagonalizes \(H(z)\), i.e. such that,
\begin{equation}
  \label{eq:Hdiagonal}
  H(z) = M^{T} D M \mbox{\ \ with \ \ } M^{T}M = I
\end{equation}
where,
\begin{equation}
  \label{eq:D}
  D = \diag\left(\xi_{1},\xi_{2},\ldots,\xi_{m},-\xi_{m+1},\ldots,
    -\xi_{p}\right)
\end{equation}
where all the $\xi_{j} > 0$ and we have assumed that the columns
of $M$ have been ordered so that all the $m$ positive eigenvalues
appear first and all the negative eigenvalues
$-\xi_{m+1},\ldots,-\xi_{p}$ appear last so that (\ref{eq:Hdiagonal})
agrees with (\ref{eq:D}).
Split $D$ as,
\begin{eqnarray}
  \label{eq:D+-}
  D &=& \diag\left(\xi_{1},\ldots,\xi_{m},0,\ldots,0\right)
       - \diag\left(0,\ldots,0,\xi_{m+1},\ldots,\xi_{p}\right) \nonumber \\
    &=& D_{+} - D_{-} 
\end{eqnarray}
and use (\ref{eq:Hdiagonal}) and (\ref{eq:D+-}) to write,
\begin{eqnarray}
  \label{eq:Hzs}
  H(z) &=& M^{T}D_{+}M - M^{T}D_{-}M \nonumber \\
       &=& \left(D_{+}^{1/2}M\right)^{T}\left(D_{+}^{1/2}M\right) 
          - \left(D_{-}^{1/2}M\right)^{T}\left(D_{-}^{1/2}M\right)
          \nonumber \\
       &=& \Sigma_{+}^{T}\Sigma_{+} - \Sigma_{-}^{T}\Sigma_{-}
\end{eqnarray}
Using (\ref{eq:Hzs}) and the substitution $y=Ax$ we obtain,
\begin{eqnarray}
  \label{eq:objfn}
  \int_{S_{A}} x^{T}H(z)x dx &=& \int_{y^{T}y \le 1} y^{T}\left(A^{-1}\right)^{T}
   \left(\Sigma_{+}^{T}\Sigma_{+} - \Sigma_{-}^{T}\Sigma_{-} \right)
   A^{-1}y dy \nonumber \\
   &=& \int_{y^{T}y\le 1} \left<\Sigma A^{-1}y,
     \Sigma A^{-1}y\right> dy
\end{eqnarray}
where,
\begin{equation}
  \label{eq:Sigma}
  \Sigma = \Sigma_{+} + \Sigma_{-} = \left(D_{+} +
    D_{-}\right)^{1/2} M
\end{equation}
and $<.,.>$ denotes the pseudo-Riemannian inner product,
\begin{equation}
  \label{eq:<,>}
  \left<x,y\right> = \sum_{i=1}^{m}x^{i}y_{i} - 
                        \sum_{i=m+1}^{p} x^{i}y_{i} 
\end{equation}
By letting $G=\diag(1,\ldots,1,-1,\ldots,-1)$ (i.e. $m$ ones followed
by $p-m$ negative ones) be the metric with the signature of $H(z)$ we
can also write (\ref{eq:<,>}) as,
\begin{equation}
  \label{eq:xGy}
  \left<x,y\right> = x^{T} G y
\end{equation}
Let,
\begin{equation}
  \label{eq:B}
  B = [b_{1}|b_{2}|\ldots |b_{p}] = \Sigma A^{-1}
\end{equation}
where $b_{1},\ldots,b_{p}$ denote the columns of $B$.
Substituting (\ref{eq:B}) into (\ref{eq:objfn}), using the linearity
of the inner product and the symmetry of the unit euclidian sphere we
obtain, 
\begin{eqnarray}
  \label{eq:objfn2}
    \int_{S_{A}} x^{T}H(z)x dx &=& \int_{y^{T}y \le 1}
        \left<By,By\right> dy  \nonumber \\
        &=& \sum_{j}\sum_{k} \left<b_{j},b_{k}\right> \int_{S_{I}}
              y^{j}y^{k} dy \\
        &=& \sum_{j}\sum_{k} \left<b_{j},b_{k}\right> \delta^{jk} \rho
                  \nonumber \\
        &=& \rho \sum_{j=1}^{p} \left<b_{j},b_{j}\right>   \label{eq:objfn3}
\end{eqnarray}
where $\rho$ stands for,
\begin{equation}
  \label{eq:rho}
  \rho = \int_{S_{I}} \left(y^{1}\right)^{2} dy = \frac{2^{p}\pi}{p+3}
\end{equation}
From (\ref{eq:B}) and (\ref{eq:objfn3}) it follows that problem
(\ref{eq:minA}) is equivalent to,
\begin{equation}
  \label{eq:minA2}
  \min_{|B| = |\Sigma|} \left(\sum_{j=1}^{p}\left<b_{j},b_{j}\right>
               \right)^{2} 
\end{equation}
When $H(z)$ is indefinite, i.e. when \(m \notin \{0,p\}\) it is
possible to choose the columns of $B$ so that 
$\sum_{j}\left<b_{j},b_{j}\right> = 0$ achieving the global
minimum. There are many possible choices but the simplest one is,
\begin{equation}
  \label{eq:Bdiag}
  B = c \cdot 
       \diag(\underbrace{\sqrt{p-m},\sqrt{p-m},\ldots,\sqrt{p-m}}_{m},
    \underbrace{\sqrt{m},\sqrt{m},\ldots,\sqrt{m}}_{p-m})
\end{equation}
since,
\begin{equation}
  \label{eq:obj=0}
  \sum_{j=1}^{p}\left<b_{j},b_{j}\right> = c^{2} m (\sqrt{p-m})^{2} - 
     c^{2} (p-m) (\sqrt{m})^{2} = 0.
\end{equation}
The scalar constant $c$ needs to be adjusted to satisfy the
constraint that $\det B = \det \Sigma$. From
(\ref{eq:B}), (\ref{eq:Bdiag}) and (\ref{eq:Sigma}) we obtain
that the matrix for the optimal metric can be written as,
\begin{equation}
  \label{eq:A2}
  A = B^{-1}\Sigma = \frac{c^{-1}}{\sqrt{p-m}}\Sigma_{+} +
        \frac{c^{-1}}{\sqrt{m}}\Sigma_{-}
\end{equation}
From (\ref{eq:A2}) we get,
\begin{equation}
  \label{eq:ATA}
  A^{T}A = \frac{c^{-2}}{p-m} \Sigma_{+}^{T}\Sigma_{+} +
     \frac{c^{-2}}{m} \Sigma_{-}^{T}\Sigma_{-}
\end{equation}
Finally from (\ref{eq:Hzs}) we can rewrite (\ref{eq:ATA}) as,
\begin{eqnarray}
  \label{eq:ATA1}
  A^{T}A &=& c^{-2} M^{T}\left( \frac{1}{p-m} D_{+} + 
           \frac{1}{m} D_{-} \right) M \\
  \label{eq:ATA2}
         &=& c^{-2} M^{T} \diag(\frac{\xi_{1}}{p-m},\ldots,
           \frac{\xi_{m}}{p-m},\frac{\xi_{m+1}}{m},\ldots,
             \frac{\xi_{p}}{m}) M 
\end{eqnarray}
Notice that when $p-m = m$ (i.e. when the number of positive equals
the number of negative eigenvalues of $H(z)$) the factor $1/m$ can be
factorized out from the diagonal matrix in (\ref{eq:ATA2}) and in this
case the optimal $A$ can be expressed as,
\begin{equation}
  \label{eq:bestA1}
  A = \frac{H_{+}^{1/2}(z)}{|H_{+}^{1/2}(z)|^{1/p}}
\end{equation}
where we have used the positive square-root of $H(z)$ defined as,
\begin{equation}
  \label{eq:H+1/2}
  H_{+}^{1/2}(z) = \diag(\sqrt{\xi_{1}},\ldots,\sqrt{\xi_{p}})M
\end{equation}
In all the other cases for which $H(z)$ is indefinite, i.e. when 
$m\notin\{0,p/2,p\}$ we have,
\begin{equation}
  \label{eq:bestA}
  A = c^{-1} \diag(\sqrt{\frac{\xi_{1}}{p-m}},\ldots,
           \sqrt{\frac{\xi_{m}}{p-m}},\sqrt{\frac{\xi_{m+1}}{m}},\ldots,
             \sqrt{\frac{\xi_{p}}{m}}) M
\end{equation}
The normalization constant $c$ is fixed by the constraint that 
$\det A = 1$ as,
\begin{equation}
  \label{eq:c}
  c = (p-m)^{-\frac{m}{2p}} m^{-\frac{(p-m)}{2p}} 
           |\det H(z)|^{\frac{1}{2p}} 
\end{equation}
This shows (\ref{eq:bestA0}).

Let us now consider the only other remaining case when $H(z)$ is
definite, i.e. either positive definite ($m=p$) or negative definite
($m=0$).  Introducing $\lambda_{0}$ as the Lagrange multiplier
associated to the constraint $\det B = \det \Sigma$ we obtain that the
problem to be solved is,
\begin{equation}
  \label{eq:minALagrange}
  \min_{b_{1},\ldots b_{p},\lambda_{0}}
           {\mathcal{L}}(b_{1},b_{2},\ldots,b_{p},\lambda_{0})
\end{equation}
where the Lagrangian \(\mathcal{L}\) is written as a function of the
columns of $B$ as,
\begin{equation}
  \label{eq:Lagrangian}
  {\mathcal{L}}(b_{1},b_{2},\ldots,b_{p},\lambda_{0}) =  
        \left(\sum_{j=1}^{p}\left<b_{j},b_{j}\right>\right)^{2} 
   - 4 \lambda_{0} (\det(b_{1},\ldots,b_{p})-\det \Sigma)
\end{equation}
The $-4\lambda_{0}$ instead of just $\lambda_{0}$ is chosen to simplify
the optimality equations below. The optimality conditions are,
\begin{equation}
  \label{eq:dLdbj}
  \pd{{\mathcal{L}}}{b_{j}} = 0  \mbox{\ \ for\ \ }  j=1,\ldots,p
    \mbox{\ \ and \ \ }\pd{{\mathcal{L}}}{\lambda_{0}} = 0
\end{equation}
where the functional partial derivatives are taken in the Fr\'{e}chet
sense with respect to the column vectors $b_{j}$. The Fr\'{e}chet
derivatives of quadratic and multi linear forms are standard text-book
exercises. Writing the derivatives as linear functions of the vector
parameter $h$ we have,
\begin{eqnarray}
  \label{eq:dbh1}
  \pd{\ }{b_{j}}\left<b_{j},b_{j}\right>(h) &=& 2\left<b_{j},h\right>
  \\   \label{eq:dbh2}
  \pd{\ }{b_{j}}\det(b_{1},\ldots,b_{p})(h) &=& 
  \det(b_{1},\ldots,\underbrace{h}_{\mbox{j-th col.}},\ldots,b_{p}) 
\end{eqnarray}
Thus, using (\ref{eq:dbh1}) and (\ref{eq:dbh2}) to compute the
derivative of (\ref{eq:Lagrangian}) we obtain that for all $h$ and all
$j=1,\ldots,p$ we must have,
\begin{equation}
  \label{eq:optL}
  \pd{{\mathcal{L}}}{b_{j}}(h) = 
      2 \left\{\sum_{k=1}^{p}\left<b_{k},b_{k}\right> \right\}
      2\left<b_{j},h\right> -
      4\lambda_{0}\det(b_{1},\ldots,h,\ldots,b_{p}) = 0
\end{equation}
When $\sum_{k}<b_{k},b_{k}> \neq 0$ we can rewrite (\ref{eq:optL})
as,
\begin{equation}
  \label{eq:<bj,h>}
  \left<b_{j},h\right> = c_{0}^{-1} \det(b_{1},\ldots,h,\ldots,b_{p})
\end{equation}
But now we can substitute $h=b_{i}$ with $i\neq j$ into
(\ref{eq:<bj,h>}) and use the fact that the determinant of a matrix
with two equal columns is zero, to obtain,
\begin{equation}
  \label{eq:<bj,bi>}
  \left<b_{j},b_{i}\right> = 0 \mbox{\ \ for  all\ \ } i \neq j.
\end{equation}
In a similar way, replacing $h=b_{j}$ into (\ref{eq:<bj,h>}), we get
\begin{equation}
  \label{eq:<bj,bj>}
  \left<b_{j},b_{j}\right> = c_{0}^{-1} \det B = c
\end{equation}
where $c$ is a constant that needs to be fixed in order to satisfy
the constraint that $\det B = \det\Sigma$. 
We have shown that the optimal matrix $B$ must have orthogonal columns
of the same length for the $G$-metric. This can be expressed with a
single matrix equation as,
\begin{equation}
  \label{eq:BTGB}
  B^{T}G B = c I
\end{equation}
Substituting (\ref{eq:B}) into (\ref{eq:BTGB}) and re-arranging terms
we obtain,
\begin{eqnarray}
  \label{eq:cATA2}
  A^{T}A &=& c^{-1} \Sigma^{T} G \Sigma \\
   &=& c^{-1} (\Sigma_{+}^{T} + \Sigma_{-}^{T}) G 
            (\Sigma_{+} + \Sigma_{-}) \nonumber \\
   &=& c^{-1} (\Sigma_{+}^{T}\Sigma_{+} - \Sigma_{-}^{T}\Sigma_{-})
            \nonumber \\
  \label{eq:cATA3}
  A^{T}A &=& c^{-1} H(z)
\end{eqnarray}
From (\ref{eq:cATA3}),~(\ref{eq:BTGB}),~(\ref{eq:rho}) and
(\ref{eq:objfn3})  we obtain,
\begin{equation}
  \label{eq:best4}
  \left| \int_{S_{A}} x^{T} H(z) x\,\,dx \right| \ge \rho p |c|
\end{equation}
and replacing the values of $\rho$ and $c$ we obtain
(\ref{eq:lb4obj}). \\
\qed

\subsection{Yet Another Proof When The Hessian is Definite}

Consider the following lemma.

\begin{lemma}
\label{lm:1}
Let $A,B$ be two $p\times p$ non-singular matrices with the same
determinant. Then
\begin{equation}
\label{eq:lemma1}
\int_{S_{A}} \|x\|^{2}_{B}  dx \ge \int_{S_{B}} \|y\|^{2}_{B} dy
\end{equation}
\end{lemma}
{\bf Proof:}
Just split $S_{A}$ and $S_{B}$ as,
\begin{eqnarray}
\label{eq:SAsplit}
S_{A} &=& (S_{A}S_{B}) \cup (S_{A}S^{c}_{B}) \\
\label{eq:SBsplit}
S_{B} &=& (S_{B}S_{A}) \cup (S_{B}S^{c}_{A})
\end{eqnarray}
and write,
\begin{equation}
\label{eq:fullsplit}
\int_{S_{A}} \|x\|^{2}_{B}  dx = \int_{S_{B}} \|x\|^{2}_{B}  dx
        - \int_{S^{c}_{A}S_{B}} \|x\|^{2}_{B}  dx +
        \int_{S_{A}S^{c}_{B}} \|x\|^{2}_{B}  dx
\end{equation}
Now clearly,
\begin{equation}
\label{eq:xinsasb}
\min_{x\in S_{A}S^{c}_{B}}\|x\|^{2}_{B} \ge 1 \ge 
        \max_{y\in S^{c}_{A}S_{B}}\|y\|^{2}_{B}
\end{equation}
from where it follows that,
\begin{eqnarray}
\label{eq:sascb1}
\int_{S_{A}S^{c}_{B}} \|x\|^{2}_{B}  dx &\ge&
        \min_{x\in S_{A}S^{c}_{B}}\|x\|^{2}_{B} 
           \int_{S_{A}S^{c}_{B}}   dx   \\
\label{eq:sascb2}
 &\ge& \max_{y\in S^{c}_{A}S_{B}}\|y\|^{2}_{B} 
        \int_{S^{c}_{A}S_{B}} dy \\
\label{eq:sascb3}
 &\ge& \int_{S^{c}_{A}S_{B}} \|y\|^{2}_{B} dy
\end{eqnarray}
where (\ref{eq:sascb1}) and (\ref{eq:sascb3}) follow from
(\ref{eq:xinsasb}). To justify the middle inequality
(\ref{eq:sascb2}) notice that from (\ref{eq:SAsplit}),
(\ref{eq:SBsplit}) and the hypothesis that $|A|=|B|$ we can write,
\begin{equation}
  \label{eq:isasbc}
  \int_{S_{A}S^{c}_{B}}   dx  +   \int_{S_{A}S_{B}}   dx =
          \int_{S^{c}_{A}S_{B}} dy + \int_{S_{A}S_{B}} dy 
\end{equation}
The conclusion (\ref{eq:lemma1}) follows from inequality
(\ref{eq:sascb3}) since that makes the last two terms in
(\ref{eq:fullsplit}) non-negative. \\
\qed 

If $B$ is a nonsingular matrix we define,
\begin{equation}
  \label{eq:Bhat}
  \hat{B} = \frac{B}{\left|\det B\right|^{1/p}}
\end{equation}
An immediate consequence of lemma \ref{lm:1} is,
\begin{theorem}
  \label{th:theth}
If $H(z)$ is definite, then for all $p\times p$ matrices with $|A|=1$
we have,
\begin{equation}
  \label{eq:corlm1}
\Delta = \int_{S_{A}} \|x\|^{2}_{H^{1/2}(z)}  dx \ge 
      \int_{S_{\hat{H}^{1/2}(z)}} \|x\|^{2}_{H^{1/2}(z)}  dx 
\end{equation}
\end{theorem}
{\bf Proof:} 
\begin{eqnarray}
  \label{eq:pfcorlm1}
  \Delta &=& 
    \left|H(z)\right|^{1/p}\int_{S_{A}}\|x\|^{2}_{\hat{H}^{1/2}(z)} dx \\
  \label{eq:pfcorlm12}
   &\ge&
  \left|H(z)\right|^{1/p}\int_{S_{\hat{H}^{1/2}(z)}}
       \|x\|^{2}_{\hat{H}^{1/2}(z)} dx \\
  \label{eq:pfcorlm13}
   &=& \int_{S_{\hat{H}^{1/2}(z)}} \|x\|^{2}_{H^{1/2}(z)}  dx 
\end{eqnarray}
where we have used lemma \ref{lm:1} to deduce the middle inequality
(\ref{eq:pfcorlm12}).\\
\qed

\subsection{Best Norm With General Kernels}
\label{sec:genKernels}
In this section we solve the problem of finding the optimal norm
in the general class of estimators (\ref{eq:doublesmooth}).

Before we optimize the norm we need to state explicitly what it means
to do estimation with different kernels and different norms. First of
all a general kernel function is a nonnegative bounded function
defined on the unit sphere generated by a given norm. Hence, the
kernel only makes sense relative to the given norm. To indicate this
dependence on the norm we write $K_{A}$ for the kernel associated to
the norm generated by the matrix $A$. We let
\begin{equation}
  \label{eq:KAK0A}
  K_{A}=K\circ A
\end{equation}
where $K=K_{I}$ is a fix mother kernel defined on the euclidian unit
sphere. Equation (\ref{eq:KAK0A}) provides meaning to the notion of
changing the norm without changing the kernel. What this means is not
that the kernel is invariant under changes of $A$ but rather
equivariant in the form specified by (\ref{eq:KAK0A}).  Recall also
that a proper kernel must satisfy (\ref{eq:intK}).  To control bias we
must also require the kernels to satisfy (\ref{eq:intxK}). It is easy
to see (just change the variables) that if the mother kernel $K$ has
these properties so do all its children $K_{A}$ with the only proviso
that $|A|=1$ in order to get (\ref{eq:intK}). Notice also that radial
symmetry of $K$ is a sufficient but not a necessary condition for
(\ref{eq:intxK}).

The optimization of the norm with general kernels looks more
complicated than when the kernel is uniform since the best
$(AMSE)^{*}$ also depends on $\int_{S_{A}}K_{A}^{2}(x)dx$. Consider the
double smoothing estimators, which are the most general case treated
in this paper. From, (\ref{eq:Bzfn}), (\ref{eq:Vzfn}) and
(\ref{eq:amse*}) we have,
\begin{equation}
  \label{eq:amse*4gen}
  (AMSE)^{*} = (\mbox{const.}) \left\{\beta^{-1}\int_{S_{A}}K_{A}^{2}(x)dx -
    \lambda_{1} \right\}^{\frac{4}{p+4}}\, f(z)\, \left(
    \frac{\Delta^{2}}{f(z)}\right)^{\frac{p}{p+4}}
\end{equation}
where the constant depends only on the dimension of the space.  Even
though the dependence of (\ref{eq:amse*4gen}) on $A$ looks much more
complicated than (\ref{eq:amse*2}) this is only apparently so.  In
fact the two expressions define very similar optimization problems as
we now show.

First notice that the search for best $A$ must be done within the
class of matrices with a fix determinant. For otherwise we will be
changing the value of the smoothness parameter that was fixed to the
best possible value in order to obtain (\ref{eq:amse*4gen}).
If we let $|A|=1$ we have,
\begin{equation}
  \label{eq:a=b=c}
  \int_{S_{A}}K(x)\, dx = \beta = \int_{S_{A}}dx = \lambda(S_{I})
\end{equation}
We also have that,
\begin{equation}
  \label{eq:K2A}
  \int_{S_{A}} K_{A}^{2}(y)\, dy =   \int_{S_{A}} K^{2}(Ay)\, dy =
      \int_{S_{I}} K^{2}(y)\, dx
\end{equation}
From (\ref{eq:a=b=c}) and (\ref{eq:K2A}) we deduce that the term in
(\ref{eq:amse*4gen}) within cursive brackets is the same for all
matrices $A$ and it depends only on the fix kernel $K$.
Finally notice that the value of $\Delta$ in (\ref{eq:amse*4gen}) is
given by
\begin{equation}
  \label{eq:gendelta}
  \Delta =  \int_{S_{A}} x^{T}H(z)x\, G(Ax)\, dx
\end{equation}
where $G(x)=K(x)+\lambda_{0}$ in the general case. By retracing again the
steps that led to (\ref{eq:objfn3}) we can write,
\begin{eqnarray}
  \label{eq:DeltaGen}
  \Delta &=& \sum_{j}\sum_{k} \left<b_{j},b_{k}\right> \int_{S_{I}}
              y^{j}y^{k} G(y)\, dy \\
        &=& \sum_{j}\sum_{k} \left<b_{j},b_{k}\right> \delta^{jk} \rho_{k}(G)
                  \nonumber \\
  \label{eq:objfngen}
        &=& \sum_{j=1}^{p} \left<b_{j},b_{j}\right> \rho_{j}(G)  
\end{eqnarray}
where now,
\begin{equation}
  \label{eq:rhoG}
  \rho_{j}(G) = \int_{S_{I}} \left(x^{j}\right)^{2} G(x)\, dx
\end{equation}
There are three cases to be considered.
\begin{enumerate}
\item All the $\rho_{j}(G)=\rho$ for $j=1,\ldots,p$. The optimization
  problem reduces to the case when the kernel is uniform and therefore
  it has the same solution.
\item All the $\rho_{j}(G)$ have the same sign, i.e. they are all
  positive or all negative. If e.g. all $\rho_{j}>0$ just replace
  $b_{j}$ with $\sqrt{\rho_{j}}b_{j}$ and use the formulas obtained
  for the uniform kernel case.
\item Some of the $\rho_{j}(G)$ are positive and some are
  negative. This case can be handled like the previous one after
  taking care of the signs for different indices $j$.
\end{enumerate}
The first case is the most important for it is the one implied when
the kernels are radially symmetric. The other two cases are only
included for completeness. Clearly if we do estimation with a non
radially symmetric kernel the optimal norm would have to correct for
this arbitrary builtin asymmetry, effectively achieving at the end the
same performance as when radially symmetric kernels are used. The
following theorem enunciates the main result.
\begin{theorem}
\label{th:genTheorem}
  In the general class of estimators (\ref{eq:doublesmooth}) with
  radially symmetric (mother) kernels, best possible asymptotic
  performance (under second order smoothness conditions) is achieved
  when distances are measured with the best metrics obtained when the
  kernel is uniform.
\end{theorem}

\section{Asymptotic Relative Efficiencies}

The practical advantage of using density estimators that adapt to the
form of the optimal metrics can be measured by computing the
Asymptotic Relative Efficiency (ARE) of the optimal metric to the
euclidian metric. Let us denote by $AMSE(I)$ and $AMSE(H(z))$ the
expressions obtained from (\ref{eq:amse*4gen}) when using the
euclidian norm and the optimal norm respectively. For the Euclidean
norm we get,
\begin{equation}
  \label{eq:AMSE(I)}
  AMSE(I) = (\mbox{const.}) \left\{\beta^{-1}\int_{S_{I}}K^{2}(x)dx -
    \lambda_{1} \right\}^{\frac{4}{p+4}}\, f(z)\, \left(
    \frac{(\rho\, \mbox{tr\ } H(z))^{2}}{f(z)}\right)^{\frac{p}{p+4}}
\end{equation}
where \mbox{tr\ } stands for the trace since,
\begin{equation}
  \label{eq:DeltaEucl}
  \Delta = \int_{S_{I}}x^{T}H(z)x\, G(x)\, dx 
   = \sum_{i,j} h_{ij}(z) \int_{S_{I}} x^{i}x^{j} G(x)\, dx
   = \rho\,\mbox{tr\ } H(z)
\end{equation}
Using (\ref{eq:a=b=c}), (\ref{eq:K2A}) and (\ref{eq:lb4obj}) we obtain
that when $H(z)$ is definite,
\begin{eqnarray}
  \label{eq:AMSE(H(z))}
\lefteqn{AMSE(H(z)) = } \\ \nonumber
 & &(\mbox{const.}) \left\{\beta^{-1}\int_{S_{I}}K^{2}(x)dx -
    \lambda_{1} \right\}^{\frac{4}{p+4}}\, f(z)\, \left(
    \frac{(\rho\, p\, |\det H(z)|^{1/p})^{2}}{f(z)}\right)^{\frac{p}{p+4}}
\end{eqnarray}
Hence, when $H(z)$ is definite the $ARE$ is,
\begin{equation}
  \label{eq:ARE}
  ARE = \frac{AMSE(I)}{AMSE(H(z))} = \left(\frac{\mbox{tr\ }H(z)}
    {p\, |\det H(z)|^{1/p}} \right)^{\frac{2p}{p+4}}
\end{equation}
If $\xi_{1},\ldots,\xi_{p}$ are the absolute value of the
eigenvalues of $H(z)$ then we can write,
\begin{equation}
  \label{eq:arith/geom}
  ARE =
 \left(\frac{\frac{1}{p}\sum_{j}\xi_{j}}
   {\left(\prod_{j}\xi_{j}\right)^{1/p}} \right)^{\frac{2p}{p+4}}
 = \left( \frac{\mbox{arith. mean of\ }\{\xi_{j}\}}
     {\mbox{geom. mean of\ }\{\xi_{j}\}} \right)^{\frac{2p}{p+4}}
\end{equation}
It can be easily shown that the arithmetic mean is always greater or
equal than the geometric mean (take logs, use the strict concavity of
the logarithm and Jensen's inequality) with equality if and only if
all the $\xi_{j}$'s are equal. Thus, it follows from
(\ref{eq:arith/geom}) that the only case in which the use of the
optimal metric will not increase the efficiency of the estimation of
the density at a point where the Hessian is definite is when all the
eigenvalues of $H(z)$ are equal. It is also worth noticing that the
efficiency increases with $p$, the dimension of the data space. There
is of course infinite relative efficiency in the regions where the
$H(z)$ is indefinite.

\section{An Example: Radially Symmetric Distributions}
When the true density $f(z)$ has radial symmetry it is possible to
find the regions where the Hessian $H(z)$ is positive and negative
definite. These models have horizons defined by the boundary between
the regions where $H(z)$ is definite. We show also that when and only
when  the density is linear in the radius of symmetry, the Hessian is
singular in the interior of a solid sphere. Thus, at the interior of
these spheres it is impossible to do estimation with the best metric.

Let us denote simply by $L$ the log likelihood, i.e.,
\begin{equation}
  \label{eq:L}
  f(z) = \exp(L)
\end{equation}
If we also denote simply by $L_{j}$ the partial derivative of $L$ with
respect to $z_{j}$ then,
\begin{equation}
  \label{eq:fj}
  \pd{f}{z_{j}} = f(z)\, L_{j}
\end{equation}
and also,
\begin{equation}
  \label{eq:fij}
  \frac{\partial^{2}f}{\partial z_{i}\partial z_{j}} =
  \pd{f}{z_{i}}\, L_{j} + f(z)\, L_{ij} = 
  f(z)\left\{ L_{i}L_{j} + L_{ij} \right\}
\end{equation}
where we have used (\ref{eq:fj}) and the definition $L_{ij} =
\pd{L_{j}}{z_{i}}$. It is worth noticing, by passing, that
(\ref{eq:fij}) implies a notable connection with the so called
nonparametric Fisher information ${\mathcal{I}}(f)$ matrix,
\begin{equation}
  \label{eq:Info}
  \int H(z)\, dz = {\mathcal{I}}(f) - {\mathcal{I}}(f) = 0
\end{equation}
our main interest here however, is the computation of the Hessian when
the density is radially symmetric. Radial symmetry about a fix point
$\mu\in \Re^{p}$ is obtained when $L$ (and thus $f$ as well) depends
on $z$ only through the norm $\|z-\mu\|_{V^{-1}}$ for some symmetric
positive definite $p\times p$ matrix $V$. Therefore we assume that,
\begin{equation}
  \label{eq:L=}
  L = L(-\frac{1}{2}(z-\mu)^{T}V^{-1}(z-\mu))
\end{equation}
from where we obtain,
\begin{eqnarray}
  \label{eq:Li}
  L_{i} &=&  \left(-v^{i\cdot} (z-\mu)\right) L^{\prime} \\
  \label{eq:Lij}
  L_{ij} &=& L^{\prime\prime} v^{i\cdot}(z-\mu)v^{j\cdot}(z-\mu)
               - L^{\prime}v^{ij}
\end{eqnarray}
where $v^{i\cdot}$ and $v^{ij}$ denote the $i$-th row and $ij$-th
entries of $V^{-1}$ respectively. Replacing (\ref{eq:Li}) and
(\ref{eq:Lij}) into (\ref{eq:fij}), using the fact that $V^{-1}$ is
symmetric and that $v^{j\cdot}(z-\mu)$ is a scalar and thus, equal to
its own transpose $(z-\mu)^{T}v^{\cdot j}$, we obtain
\begin{equation}
  \label{eq:HzfzL}
  H(z) = f(z) L^{\prime} \left\{
    \left( L^{\prime}+\frac{L^{\prime\prime}}{L^{\prime}}\right)
    V^{-1}(z-\mu)(z-\mu)^{T} - I \right\} V^{-1}
\end{equation}
We have also assumed that $L^{\prime}$ is never zero. With the help of
(\ref{eq:HzfzL}) we can now find the conditions for $H(z)$ to be
definite and singular. Clearly $H(z)$ will be singular when the
determinant of the matrix within curly brackets in (\ref{eq:HzfzL}) is
zero. But that determinant being zero means that $\lambda=1$ is an
eigenvalue of
\begin{equation}
  \label{eq:Lprime}
    (L^{\prime}+L^{\prime\prime}/L^{\prime})V^{-1}(z-\mu)(z-\mu)^{T}
\end{equation}
and since this last matrix has rank one its only nonzero eigenvalue
must be equal to its own trace. Using the cyclical property of the
trace and letting 
\[y=-\frac{1}{2}(z-\mu)^{T}V^{-1}(z-\mu)\]
we can write,
\begin{theorem}
  \label{th:Hradsym}
  The Hessian of a radially symmetric density is singular when and
  only when either $L^{\prime}=0$ or
  \begin{equation}
    \label{eq:LLL}
    L^{\prime}+ \frac{d}{dy}\log L^{\prime} = -\frac{1}{2y}
  \end{equation}
\end{theorem}

Notice that theorem \ref{th:Hradsym} provides an equation in $y$ after
replacing a particular function $L=L(y)$. Theorem \ref{th:Hradsym} can
also be used to find the functions $L(y)$ that will make the Hessian
singular. Integrating (\ref{eq:LLL}) we obtain,
\begin{equation}
  \label{eq:Ly+log}
  L(y) + \log L^{\prime}(y) = -\frac{1}{2}\log(|y|) + c
\end{equation}
and solving for $L^{\prime}$, separating the variables and integrating
we get,
\begin{equation}
  \label{eq:Ly=log}
  L(y) = \log \left( a \sqrt{|y|} + b \right)
\end{equation}
where $a$ and $b$ are constants of integration.
In terms of the density equation (\ref{eq:Ly=log}) translates to,
\begin{equation}
  \label{eq:fzLy}
  f(z) = a\|z-\mu\|_{V^{-1}}+b
\end{equation}
Hence, in the regions where the density is a straight line as a
function of $r=\|z-\mu\|_{V^{-1}}$ the Hessian is singular and
estimation with best metrics is not possible.
Moreover, from (\ref{eq:HzfzL}) we can also obtain the regions of
space where the Hessian is positive and where it is negative
definite. When $L^{\prime}>0$, $H(z)$ will be negative definite
provided that the matrix,
\begin{equation}
  \label{eq:I-V}
 I - (L^{\prime}+L^{\prime\prime}/L^{\prime})V^{-1}(z-\mu)(z-\mu)^{T}
\end{equation}
is positive definite. But a matrix is positive definite when and only
when all its eigenvalues are positive. It is immediate to verify that
$\xi$ is an eigenvalue for the matrix (\ref{eq:I-V}) if and only if
$(1-\xi)$ is an eigenvalue of the matrix (\ref{eq:Lprime}). The matrix
(\ref{eq:Lprime}) has rank one and therefore its only nonzero
eigenvalue is its trace so we arrive to,
\begin{theorem}
  \label{th:H+-neg}
  When,
  \begin{equation}
    \label{eq:H-def}
    L^{\prime}+ \frac{d}{dy}\log L^{\prime} < -\frac{1}{2y}    
  \end{equation}
$H(z)$ is negative definite when $L^{\prime}>0$ and positive
definite when $L^{\prime}<0$. When,
\begin{equation}
  \label{eq:H+def}
    L^{\prime}+ \frac{d}{dy}\log L^{\prime} > -\frac{1}{2y}    
\end{equation}
$H(z)$ is indefinite.
\end{theorem}

For example when $f(z)$ is multivariate Gaussian $L(y)=y+c$ so that
$L^{\prime}=1$ and the horizon is the surface of the $V^{-1}$-sphere
of radius one i.e., $(z-\mu)^{T}V^{-1}(z-\mu)=1$. Inside this sphere
the Hessian is negative definite and outside the sphere the Hessian is
indefinite. The results in this section can be applied to any other
class of radially symmetric distributions, e.g. multivariate $T$ which
includes the Cauchy.

\section{Conclusions}
We have shown the existence of optimal metrics in nonparametric
density estimation. The metrics are generated by the Hessian of the
underlying density and they are unique in the regions where the
Hessian is definite. The optimal metric can be expressed as a
continuous function of the Hessian in the regions where it is
indefinite. The Hessian varies continuously from point to point thus,
associated to the general class of density estimators
(\ref{eq:doublesmooth}) there is a Riemannian manifold with the
property that if the estimators are computed based on its metric the
best asymptotic mean square error is minimized. The results are
sufficiently general to show that these are absolute bounds on the
quality of statistical inference from data. 

The similarities with General Relativity are evident but so are the
differences. For example, since the Hessian of the underlying density
is negative definite at local maxima, it follows that there will be a
horizon boundary where the Hessian becomes singular. The cross of the
boundary corresponds to a change of signature in the metric. These
horizons almost always are null sets and therefore irrelevant from a
probabilistic point of view. However, when the density is radially
symmetric changing linearly with the radius we get solid spots of
singularity. There is a qualitative change in the quality of inference
that can be achieved within these dark spots. But unlike GR, not only
around local maxima but also around local minima of the density we
find horizons. Besides, it is not necessary for the density to reach a
certain threshold for these horizons to appear. Nevertheless, I
believe that the infusion of new statistical ideas into the
foundations of Physics, specially at this point in history, should be
embraced with optimism. Only new data will (help to) tell.

There are many unexplored promising avenues along the lines of the
subject of this paper but one that is obvious from a GR point of
view. What is missing is the connection between curvature and
probability density, i.e. the field equation. I hope to be able to
work on this in the near future.

The existence of optimal metrics in density estimation is not only of
theoretical importance but of significant practical value as well. By
estimating the Hessian (e.g. with kernels that can take positive and
negative values, see \cite{singh76}) we can build  estimators that
adapt to the form of the optimal norm with efficiency gains that
increase with the number of dimensions. The antidote to the curse of
dimensionality!

\section{Acknowledgments}
I would like to thank my friends in the Maximum Entropy community
specially Gary Erickson for providing a stimulating environment
for this meeting. I am also in debt to Ariel Caticha for many
interesting conversations about life, the universe, and these things.

\bibliography{carlos} \bibliographystyle{maxent95}

\begin{thebibliography}{10}

\bibitem{vapnik98}
V.~N. Vapnik, {\em Statistical Learning Theory}, John Wiley \& Sons, Inc.,
  1998.

\bibitem{devroye-gyorfi-lugosi96}
L.~G. L.~Devroye and G.~Lugosi, {\em A Probabilistic Theory of Pattern
  Recognition}, Springer, New York, 1996.

\bibitem{devroye-gyorfi90}
L.~Devroye and L.~Gy\"{o}rfi, ``No empirical probability measure can converge
  in the total variation sense for all distributions,'' {\em Annals of
  Statistics}, {\bf 18}, (3), pp.~1496--1499, 1990.

\bibitem{renyi59}
A.~R\'{e}nyi, ``On the dimension and entropy of probability distributions,''
  {\em Acta Math. Acad. Sci. Hungar.}, {\bf 10}, pp.~193--215, 1959.

\bibitem{jaynes57}
E.~Jaynes, ``Information theory and statistical mechanics,'' {\em Phys. Rev.},
  {\bf 106}, p.~620, 1957.
\newblock Part II; ibid, vol 108,171.

\bibitem{frieden98}
B.~R. Frieden, {\em Physics from Fisher Information, a Unification}, Cambridge
  University Press, 1998.

\bibitem{rodriguez98b}
C.~C. Rodr\'{\i}guez, ``Are we cruising a hypothesis space?,'' in {\em Maximum
  Entropy and Bayesian Methods},  R.~F. W.~von~der Linden, V.~Dose and
  R.~Preuss, eds., vol.~18, (Netherlands), pp.~131--140, Kluwer Academic
  Publishers, 1998.
\newblock Also at xxx.lanl.gov/abs/physics/9808009.

\bibitem{caticha00}
A.~Caticha, ``Change, time and information geometry,'' in {\em Maximum Entropy
  and Bayesian Methods},  A.~Mohammad-Djafari, ed., vol.~19, Kluwer Academic
  Publishers, 2000.
\newblock too appear. Also at math-ph/0008018.

\bibitem{rodriguez86}
C.~C. Rodriguez, ``On a new class of multivariate density estimators,'' tech.
  rep., Dept. of Mathematics and Statistics, The University at Albany, 1986.
\newblock (http://omega.albany.edu:8008/npde.ps).

\bibitem{rodriguez88}
C.~C. Rodriguez, ``The riemannian manifold induced by a density estimator,''
  tech. rep., Dept. of Mathematics and Statistics, The University at Albany,
  1988.
\newblock (http://omega.albany.edu:8008/rmide.html).

\bibitem{fukunaga-hostetler72}
K.~Fukunaga and L.~D. Hostetler, ``Optimization of k-nearest-neighbor density
  estimates,'' {\em IEEE Trans. on Information Theory}, {\bf IT-19},
  pp.~320--326, May 1972.

\bibitem{short-fukunaga81}
R.~D. Short and K.~Fukunaga, ``The optimal distance measure for nearest
  neighbor classification,'' {\em IEEE Trans. on Information Theory}, {\bf
  IT-27}, pp.~622--637, September 1981.

\bibitem{fukunaga-flick84}
K.~Fukunaga and T.~Flick, ``An optimal global nearest neigbor metric,'' {\em
  IEEE Trans. on Pattern Analysis and Machine Intelligence}, {\bf PAMI-6},
  pp.~314--318, May 1984.

\bibitem{fukunaga-hummels87}
K.~Fukunaga and D.~M. Hummels, ``Bayes error estimation using parzen and k-nn
  procedures,'' {\em IEEE Trans. on Pattern Analysis and Machine Intelligence},
  {\bf PAMI-9}, pp.~634--643, September 1987.

\bibitem{myles-hand90}
J.~P. Myles and D.~J. Hand, ``The multi-class metric problem in nearest
  neighbour discrimination rules,'' {\em Pattern Recognition}, {\bf 23}, (11),
  pp.~1291--1297, 1990.

\bibitem{fix-hodges51}
E.~Fix and J.~L. Hodges, ``Discriminatory analysis. nonparametric
  discrimination: Consistency properties,'' Tech. Rep. 4 Project number
  21-49-004, USAF School of Aviation Medicine, Randolph Field, Tx., 1951.

\bibitem{lecam-yang90}
L.~M. Le-Cam and G.~Lo-Yang, {\em Asymptotics in Statistics: Some Basic
  Concepts}, Springer series in statistics, Springer-Verlag, 1990.

\bibitem{diaconnis-freedman86}
P.~Diaconnis and D.~Freedman, ``On the consistency of bayesian estimates (with
  discussions),'' {\em Ann. Stat.}, {\bf 14}, (1), pp.~1--67, 1986.

\bibitem{rodriguez97}
C.~C. Rodr\'{\i}guez, ``Cv-np bayesianism by mcmc,'' in {\em Maximum Entropy
  and Bayesian Methods},  G.~J. Erickson, ed., vol.~17, Kluwer Academic
  Publishers, 1997.
\newblock (physics/9712041).

\bibitem{kn:brgmv.81}
I.~Ibragimov and R.~Has'minskii, {\em Statistical Estimation}, vol.~16 of {\em
  Applications of Mathematics}, Springer-Verlag, 1981.

\bibitem{singh76}
R.~S. Singh, ``Nonparametric estimation of mixed partial derivatives of a
  multivariate density,'' {\em Journal of Multivariate Analysis}, {\bf 6},
  pp.~111--122, 1976.

\end{thebibliography}

\end{document}